# Refined Ephemeris for Four Hot Jupiters using Ground-Based and TESS Observations

F. Davoudi[1,2,3], P. MirshafieKhozani[1,3], E. Paki[1,3], M. Roshana[1,3], F. Hasheminasab[3], A. MazidabadiFarahani[1,3], F. Ahangarani Farahani[1,3], T. Farjadnia[3], F. Nasrollahzadeh[1,3], S. Rezvanpanah[3], S.M. Mousavi[3], R. Foroughi[3], A. Poro[1,2,3], A. Ghalee[3,4]

[1]The International Occultation Timing Association Middle East section, Iran, info@iota-me.com
[2]Astronomy Department of the Raderon Lab., Burnaby, BC, Canada
[3]The Eight IOTA/ME Summer School of Astronomy, Tafresh University, Tafresh, Iran
[4]Department of Physics, Tafresh University, P.O. Box 39518-79611, Tafresh, Iran

**Abstract**

WASP-12 b, WASP-33 b, WASP-36 b, and WASP-46 b are four transiting planetary systems which we have studied. These systems' light curves were derived from observations made by the Transiting Light Exoplanet Survey Satellite (TESS) and some ground-based telescopes. We used Exofast-v1 to model these light curves and calculate mid-transit times. Also, we plotted TTV diagrams for them using derived mid-transit times and those available within the literature. O-C analysis of these timings enables us to refine the linear ephemeris of four systems. We measured WASP-12's tidal quality factor based on adding TESS data as $Q'_* = (2.13 \pm 0.29) \times 10^5$. According to the analysis, the orbital period of the WASP-46 b system is increasing. The WASP-36 b and WASP-33 b systems have not shown any obvious quadratic trend in their TTV diagrams. The increase in their period is most likely due to inaccurate liner ephemeris that has increased over time. So, more observations are needed to evaluate whether or not there is an orbital decay in the WASP-36 b and WASP-33 b systems.

## 1. INTRODUCTION

Many transiting exoplanets have been discovered to date, and such discoveries continue. Therefore, examining their characteristics and evolution has become an important issue. The ephemeris shows the place of the exoplanet in its orbit, which depends on the orbital period. Orbital period variation can be detected through Transit Timing Variation (TTV), a valid and prevalent tool for discussing these variations' apparent or physical causes. One of these physical effects is orbital decay. The orbital decay of hot Jupiters due to tidal energy dissipations could provoke their short-period orbit to shrink and eventually collide these exoplanets with their host star (Southworth et al. 2019).

The use of photometric data plays an important role in studies of target systems. Some space missions, like TESS, do photometry of transiting exoplanets. TESS is a mission launched by NASA in 2018 to search and observe planets transiting bright stars in the local neighborhood, which are 10 to 100 times brighter than Kepler mission stars (Ricker et al. 2014). Also, ground-based observations by amateur observers are also important. For instance, the Exoplanet Transit Database (ETD) is a user-friendly online portal preserved by the Czech Astronomical Society (CAS). It has been developed to concentrate and classify all available photometric data and light curves of transit observations (Brát et al. 2010, Poddaný et al. 2010).

This study investigated the ephemeris and TTVs of four hot Jupiter exoplanets with an orbital size of less than 0.027 AU and short periods of less than 1.6 days. WASP-12 b (Hebb et al. 2009), WASP-33 b (Cameron et al. 2010), WASP-36 b (Smith et al. 2012), and WASP-46 b (Anderson et al. 2012) that their host stars are solar-type. These planets were discovered by the transit method during the Wide Angle Search for Planets (WASP) project; WASP is an international project aimed at discovering exoplanets using transit photometry (Pollacco et al. 2006). We have thoroughly reviewed the history of these systems' studies. Because future observations are dependent on the previous results of the period's rate, the system's ephemeris must be refined. All the mid-transit times from discovery until 2021 have been used to update the system's ephemeris and period studies; including TESS light curves, ETD light curves, the data from the TRAPPIST, and EulerCam observations for WASP-36 b.

In the Light curve extraction section, we have discussed how we extracted the ground-based and TESS observation light curves that we used in our analysis. In the Method of analysis section, we have explained our analysis procedure in detail. In the section on period variation, we discussed the theory of period changes and examined the ephemeris changes and periods of each system through TTV diagram in four subsections.

## 2. LIGHT CURVE EXTRACTION

All the targets' host stars were observed by TESS at 2-minute cadences. Observations of WASP-12 b (TESS ID: 86396382) were obtained in sector 20 for 27 days; WASP-33 b (TESS ID: 129979528) was observed in sector 18 for 27 days, and WASP-36 b (TESS ID: 13349647) observations were made in sector 8 and will be observed in sector 34; this is included in one of the 54-day goals of TESS. A 54-day observation of WASP-46 b (TESS ID: 231663901) was obtained in sectors 1 and 27.

TESS data is available at the Mikulski Space Telescope Archive (MAST). We extracted TESS style curves using the LightKurve code[1] from MAST. LightKurve is a Python package designed to simplify the analysis of astronomical time series data collected by NASA telescopes, particularly the extrasolar planet and TESS missions. It has good practical functions to analyze the luminosity time series of star images. A light curve shows flux over time as an array of objects (Cardoso et al. 2018).

The LightKurve package includes TPFs data type, Target Pixel Files (TPFs) as the rawest form of data. TPF of WASP-12 b is shown in Figure 1. After instrumental artifacts removal, Light curves were generated, showing flux as a function of time in Barycentric TESS Julian Day (BTJD). Then we plotted them in phase. The folded transit light curve of WASP-12 b is shown in Figure 2. The folded light curves of other systems are available in Appendix.

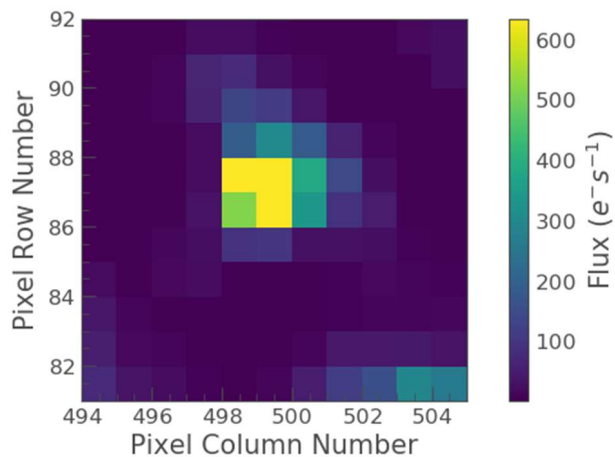

**Fig 1.** WASP-12 b (TIC 86396382) Target Pixel File, sector 20.

---

[1]https://docs.lightkurve.org/

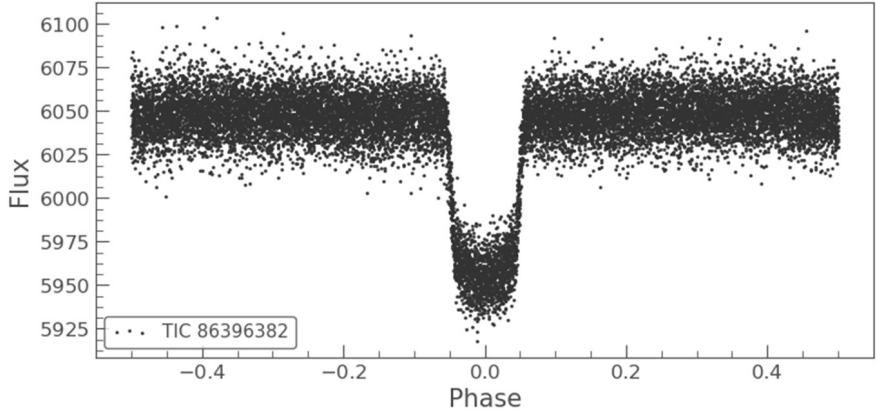

**Fig 2.** WASP-12 b folded light curve obtained from the LightKurve code.

We collected and analyzed the transit light curves obtained by amateur and professional observers publicly available through ETD (Davoudi et al. 2020). They were observed by ground-based telescopes with apertures smaller than one meter. Light curves were obtained from various filters and time scales. We examined ETD light curves with a Data Quality (DQ) index of less than DQ-3 and then removed those that had more noise or their time column reported with less decimal accuracy, and the best ones were selected.

The observations of WASP-36 b were carried out by a 0.6-meter telescope (TRAPPIST) and with a 1.2-meter telescope (Euler) at the ESO La Silla Observatory, Chile. Nine new transit light curves of WASP-36 b were obtained in 2011-2012 with TRAPPIST, and one additional transit was observed with EulerCam in 2013. We used the unpublished data that was given by these projects for this study.

## 3. METHOD OF ANALYSIS

In some ETD light curves, the airmass effect has been ignored, so airmass must be calculated based on the observers' location, which influences and improves the measured mid-transit times of related light curves. Accordingly, we computed the airmass using the Astropy package in Python (Robitaille et al. 2013).

A reduction in the star's flux occurs when its light passes through the Earth's atmosphere. This flux reduction is called the airmass and is defined by $X$

$$X = sec\theta_Z - \Delta X \qquad (1)$$

where $sec\theta_Z$ is related to the position of the object.

$$sec\theta_Z = \frac{1}{[sin\varphi sin\sigma + cos\varphi cos\sigma cos\ ]} \qquad (2)$$

where $\varphi$ represents the latitude of the observation site, $\sigma$ is the declination of the object in the sky, and $h$ is the hour angle of the object during the observation. A correction factor $\Delta X$ is subtracted from $sec\theta_Z$, due to the spherical atmosphere.

$$\Delta X = 0.00186(sec\theta_Z - 1) + 0.002875(sec\theta_Z - 1)^2 + 0.0008083(sec\theta_Z - 1)^3 \qquad (3)$$

Also, we converted the times from $HJD_{UTC}$ and $JD_{UTC}$ to the $BJD_{TDB}$ time scale (Eastman et al. 2010). For airmass-detrending on ETD light curves, we used AstroImageJ (Collins et al. 2017a), then we converted magnitudes to fluxes and normalized the fluxes to 1. The flux normalization for TESS light curves was done with the same AstroImageJ software. We obtained uniform ones in the same conditions for the purpose of light curve modeling. We applied Exofast-v1 (Eastman et al. 2013) on all light curves for modeling, and as a result, we used the output mid-transit times and their uncertainties. Figure 3 and Figure 4 show two modeled light curves of WASP-12 b. We used them for plotting TTV diagrams mentioned in the period variation section.

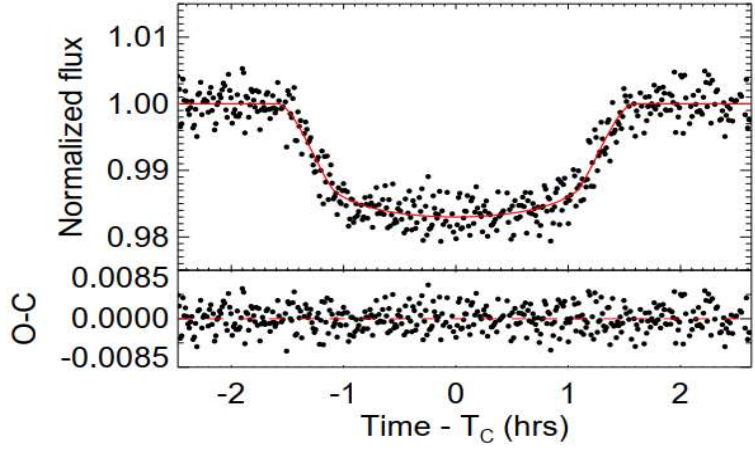

**Fig 3.** WASP-12 b ETD modeled light curve.

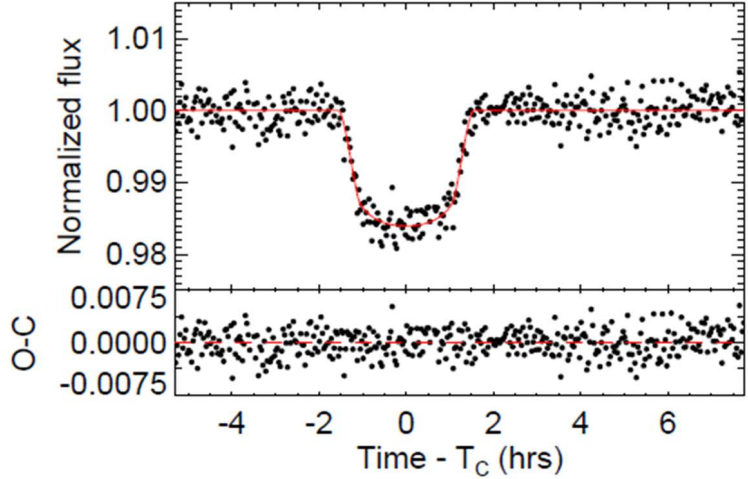

**Fig 4.** WASP-12 b modeled TESS light curve.

## 4. PERIOD VARIATIONS

The ephemeris shows the place of the exoplanet in its orbit. The general ephemeris formula is

$$T_c = T_0 + E \times P \qquad (4)$$

with $T_0$ as the timing zero point (the reference mid-transit time), $P$ as the orbital period, and $E$ as the number of epochs passed since $T_0$. The predicted mid-transit times of the exoplanet, $T_c$, can be calculated using this linear ephemeris when passing in front of its host star (Equation 4). Usually, the mid-transit time obtained from observations, $T_o$, does not have the same values as $T_c$ for the same epoch. The difference between the observed transit times and the calculated ones is known as O-C (Equation 5),

$$O - C = T_o - T_c = (\Delta T_0 + \Delta P \times E) + Q \times E^2 + \delta T_i \qquad (5).$$

$\Delta T_0$, $\Delta P$, and $Q$ are the linear change of the reference mid-transit time, the linear change of the reference period, and the quadratic change of the orbital period after passing the $E$ cycle number of the period respectively. This equation consists of three parts; The first term deals with the linear changes caused by a wrong linear ephemeris that indicates the timing uncertainty increases linearly with the number of transit epochs that have passed since $T_0$. The second term is quadratic, which describes changes due to orbital decay in planetary systems, and the last term, $\delta T_i$, means more complex periodic variations (Gajdoš & Parimucha 2019).

**4.1 WASP-12 b**

The planet WASP-12 b is a hot Jupiter studied since its discovery in 2008 due to a deviation from a linear ephemeris reported in follow-up observations. Hebb et al. (2009) reported on the planet and its host stars, characteristics obtained from the transit light curve and radial velocity. The parameters of WASP-12 b were refined in 2011, and it was speculated that the TTV signal was induced by an additional planet in this system (Maciejewski et al. 2011). Following that, surveying of the planet's mid-transit times continued, indicating that its orbital period was decreasing. Various scenarios have been proposed to describe this phenomenon; It could be caused by a perturbing planet in the system (Maciejewski et al. 2013) or may be interpreted as the result of orbital decay driven by tidal dissipation in the host star (Maciejewski et al. 2016). Patra et al. (2017) found a rate of $\dot{p} = -29 \pm 3 \frac{ms}{yr}$ and they concluded that it is difficult to distinguish that these variations indicate an orbital decay or a portion of a 14-year apsidal precession cycle. Maciejewski et al. (2018) computed $Q'_* = (1.82 \pm 0.32) \times 10^5$ as host star's tidal quality factor. Maciejewski et al. (2020) found a 5.8 sigma level for apparent eccentricity of WASP-12 b's orbit, the fact that this value is non-zero is a sign for the presence of the tidal fluid flow. However, it is also possible that the planet's orbit is slightly eccentric and is undergoing apsidal precession (Yee et al. 2019; they measured $Q'_* = 1.75^{+0.13}_{-0.11} \times 10^5$).

We performed our analysis on the ETD light curve, TESS data, and the previous literature of WASP-12 b to prepare the WASP-12 b TTV diagram (Figure 5). A supplement to this paper provides 4 machine-readable tables that contain the details of the described TTV diagram of WASP-12 b, WASP-33 b, WASP-36 b, and WASP-46 b. These tables include the mid-transit times in BJD$_{TDB}$, their errors, epochs, O-C (day), and their light curve source.

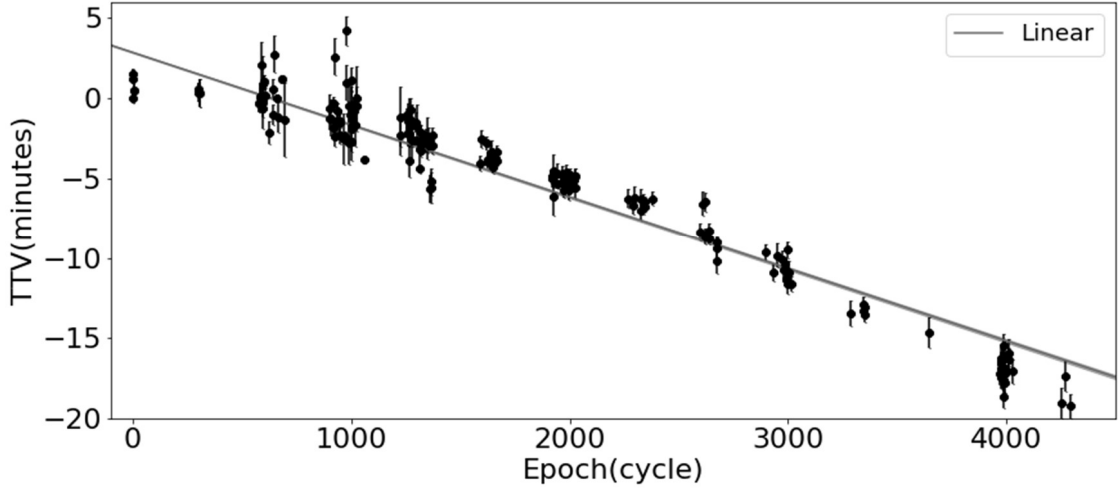

**Fig 5.** The TTV Diagram of WASP-12 b with the linear fit on the data points using the MCMC, BIC=18.55.

We used the linear ephemeris of Hebb et al. (2009) for computing epochs and the TTV values:

Reference ephemeris (BJD$_{TDB}$) = $(2454508.9761 \pm 0.0002) + (1.091423 \pm 0.000003) \times E$     (6)

We determined a new linear ephemeris by fitting a line on the TTV diagram using MCMC as,

New Ephemeris (BJD$_{TDB}$) = $(2454508.978086 \pm 0.000025) + (1.09141987 \pm 0.00000001) \times E$     (7)

For this work, we applied maximum likelihood and used the MCMC method for sampling from the posterior probability distributions of the coefficients of both models (50 walkers, 10000 step number, and 500 burn-in) using the Pymc3 package in Python (Salvatier et al. 2016). The two-dimensional probability distributions of dT and dP and the histograms of the posterior probability distribution of both obtained from MCMC have been displayed in the corner plot (Figure 6).

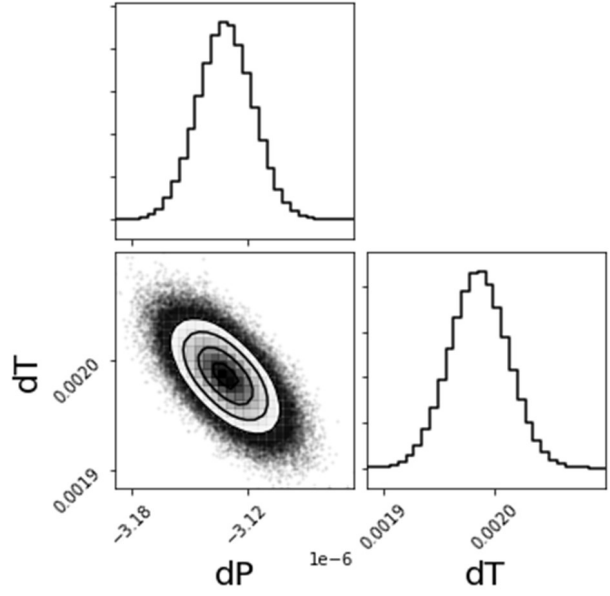

**Fig 6.** Posterior distributions for the fitted parameters using MCMC (dT & dP).

The tidal effect seems like a continuous decrease in the exoplanet's orbital period and it causes a quadratic variation in the transit times (Southworth et al. 2019)

$$T_{mid} = T_0 + PE + \frac{1}{2}P\frac{dP}{dt}E^2 \qquad (8)$$

$$\frac{dP}{dt} = \frac{-27\pi}{2Q'_*}\left(\frac{M_p}{M_*}\right)\left(\frac{R_*}{a}\right)^5 \qquad (9)$$

that $Q'_*$ is the tidal quality factor, $M_p$ and $M_*$ are the planetary and stellar masses; $R_*$ is the stellar radius and $a$ is the semi-major axis of the orbit (Patra et al. 2020).

So, to examine this system's orbital decay, we have done a quadratic fit on the residuals of the linear fit using the MCMC method (50 walkers, 10000 step numbers, and 500 burn-in), which is shown in Figure 7. And posterior distributions for the fitted parameters of the quadratic model are shown in Figure 8.
We calculated the Bayesian Information Criterion ($BIC$) (Neath & Cavanaugh 2012) where $n$ is the number of observations and $k$ is the number of model parameters, under the assumption of normality, for our models by

$$BIC = \chi^2 + k\ln(n) \qquad (10)$$

for linear fit $BIC = 18.55$ and for quadratic fit $BIC = 16.47$. The BIC value favors the quadratic model over the linear model.

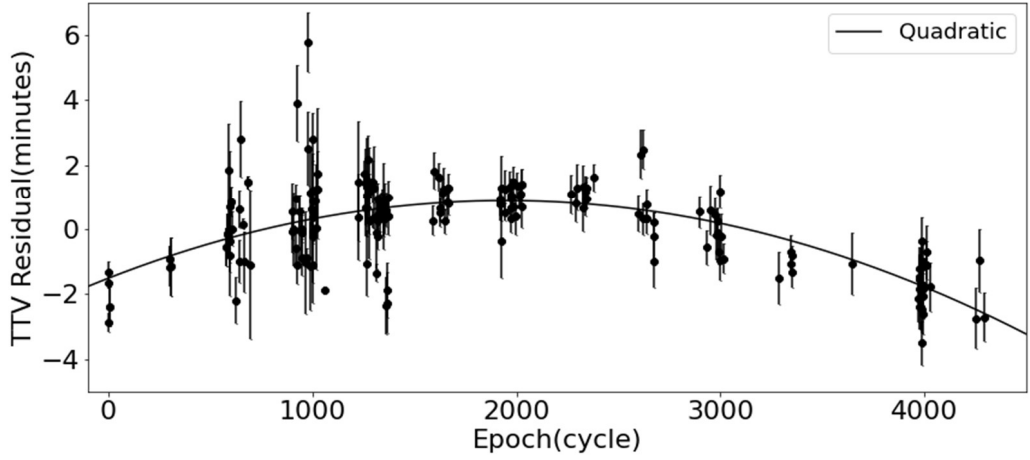

**Fig 7.** TTV Diagram of WASP-12 b with the quadratic fit on the Residuals of Linear fit, $BIC = 16.47$.

As the tidal effect causes a quadratic variation in the transit times using equations 9 and 10. Also, we assumed $M_P(M_J) = 1.465 \pm 0.079$, $M_*(M_\odot) = 1.434 \pm 0.110$, $R_*(R_\odot) = 1.657 \pm 0.046$ and $a(AU) = 0.02320 \pm 0.00064$ (Chakrabarty & Sengupta 2019). Then we found a rate of $\frac{dP}{dt} = (-8.03 \pm 0.45) \times 10^{-10}$ for this system from equation 9. Using $\frac{dP}{dt}$ and equation 9 and employing the error propagation for finding the error value, we calculated $Q'_* = (2.13 \pm 0.29) \times 10^5$ for the host star's tidal quality parameter.

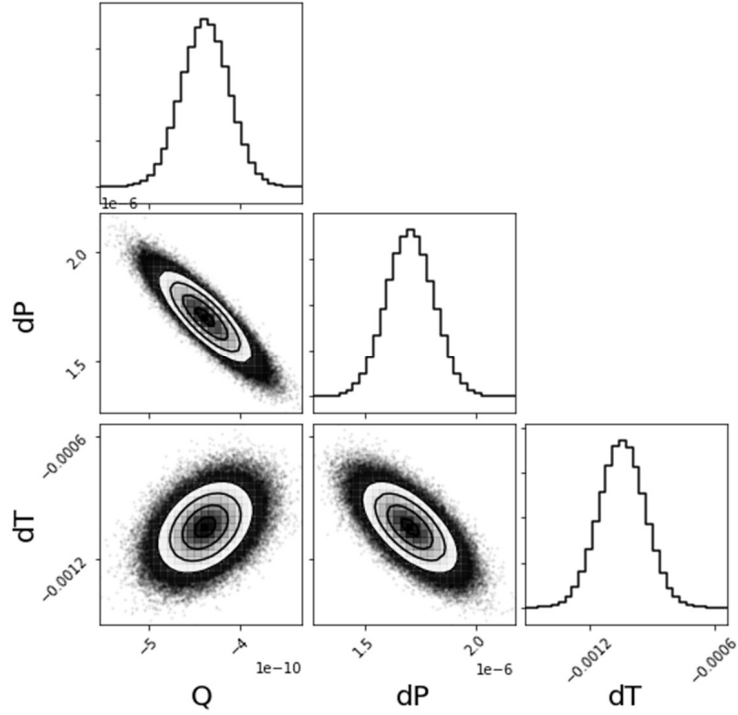

**Fig 8.** Posterior distributions for the fitted parameters of the quadratic model using MCMC. Q is the quadratic coefficient of the model.

**4.2 WASP-46 b**

WASP-46 b is a massive hot Jupiter discovered by Anderson et al. (2012). Weak Ca II (H and K) emission in the host star's spectra shows an active photosphere and chromosphere. Unseen stellar spots induced anomalies during the transit of some light curves. Ciceri et al. (2016) performed the first transit timing variation (TTV) study for this exoplanet. Their results indicate that the linear ephemeris is not an appropriate model for the observations. Also, they did not find any periodic trend in the TTV diagram. However, they pointed out that a homogeneous TTV diagram analysis using more precise data is needed.

Petrucci t al. (2018) collected new transit light curves of WASP-46 b from three different telescopes and studied homogeneous TTVs from analysis of their new and existing literature data. They investigated the correlation between O-C values, correlated noise level (β), and the Photon Noise Rate (PNR), then performed a linear fit on the data. β and PNR point out the red and white noise of light curves respectively (Winn et al. 2008, Fulton et al. 2011). Therefore, they recognized that points with less than 1-minute uncertainty and no anomalies are the best points for drawing the system's TTV diagram due to the star's magnetic activity. Petrucci et al. (2018) suggested that there are high scattering data points in the TTV diagram. However, they also did not observe any periodic trends. They also searched for possible orbital decay and obtained $Q > 7 \times 10^3$.

We have used table seven-Petrucci et al. (2018) for all mid-transit times before epoch=1552. We analyzed four high-quality ETD light curves after epoch=1551, along with TESS light curves, and obtained the mid-transit times and the uncertainties. We have continued their method by choosing data points with uncertainties of less than 1 min, according to the results of Petrucci et al. (2018). We plotted a new TTV diagram (Figure 9) using the ephemeris given by Petrucci et al. (2018) (Equation 11) and realized that a decrease in the orbital period does not continue, and TTV values become positive. We checked the TTV diagram with all TESS data with any uncertainties; the result is the same, and the period is increasing, so the orbital decay model has become weak in this system. As suggested by previous papers, this trend could be caused by star magnetic activity.

$$\text{Reference ephemeris (BJD}_{\text{TDB}}) = 2455392.3176(2) + 1.43037123(26) \times E \quad (11)$$

Then we derived the new ephemeris by a linear fit of the TTV diagram as

$$\text{New Ephemeris (BJD}_{\text{TDB}}) = (2455392.314297_{-0.000043}^{+0.000042}) + (1.43037276 \pm 0.00000004) \times E \quad (12)$$

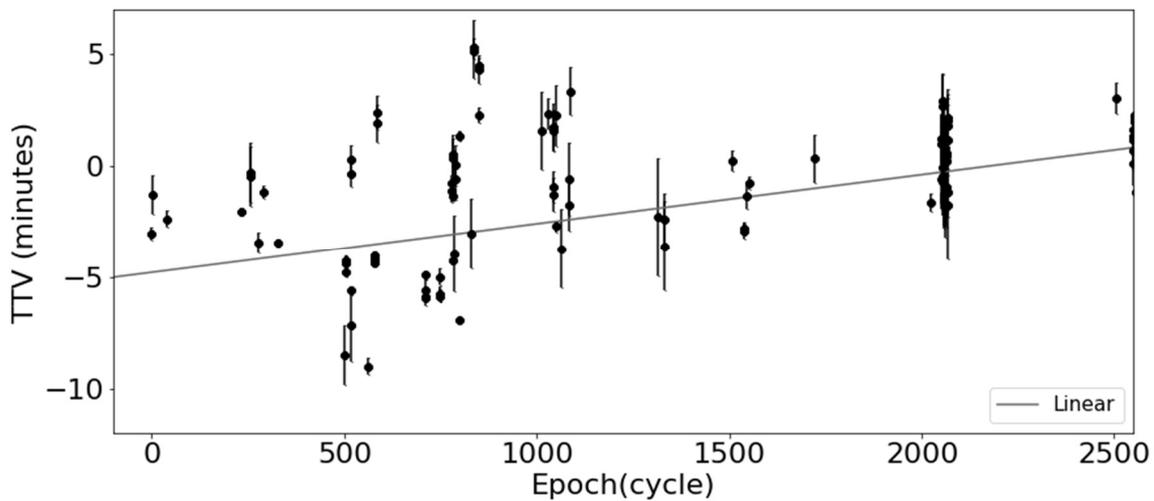

**Fig 9.** The TTV Diagram of WASP-46 b with the linear fit on the data points using the MCMC.

The corner plot of dp and dT obtained from MCMC was displayed in Figure 10. We applied 50 walkers, 10000 step numbers, and 500 burn-in for MCMC sampling.

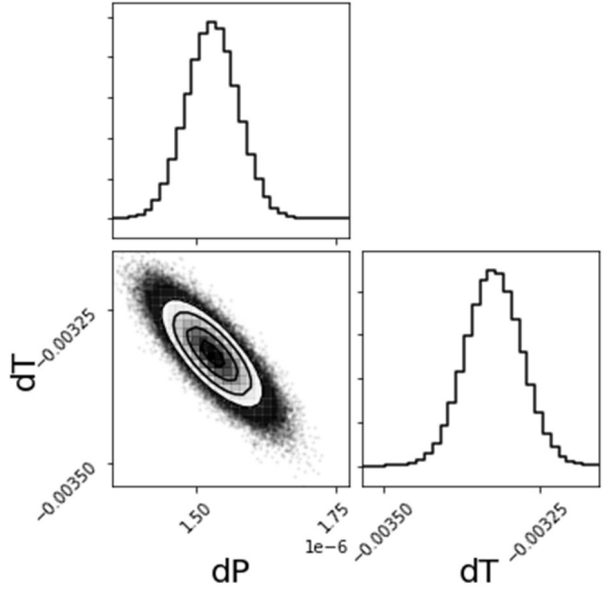

**Fig 10.** Posterior distributions for the fitted parameters using MCMC (dT & dP).

**4.3 WASP-36 b**

WASP-36 b (Smith et al. 2012), a planet with a 2.30 Jupiter-Mass, is a hot Jupiter orbiting around a metal-poor sun-like star. The initial observations of this planet have been made by Smith et al. (2012) to determine the physical properties of the system. They found no significant deviation from the reference mid-transit time. Radial velocity measurements by Smith et al. (2012) indicated no signs of an additional body in the system. Photometric observation analysis by them, due to the search for star-spots or stellar rotation effect revealed no evidence. Even the Ca II K+H emission absenteeism in star spectra in the spectroscopic observations proved the lack of stellar activity. Despite the size of WASP-36 b in close orbit, Smith et al. (2012) found no tidal effect in the system. Zhou et al. (2015) presented a secondary eclipse measurement of WASP-36 b and declared that no clear eclipse was seen in the observations. In 2016, Mancini et al. updated the reference ephemeris and stated that linear fitting is not an appropriate trend for the TTV diagram's data points. Essick & Weinberg (2015) measured non-linear orbital decay of hot Jupiters around sun-like stars based on the star's tidal dissipation. They predicted that the orbital decay of WASP-36 b in 10 years would be around P[ms] = 4.4 $\pm$ 0.15, with a minimum of P[ms] = 2.3 and a maximum of P[ms] = 9.3.

We used the mid-transit times for preparing the TTV diagram, resulting from the modeling of ETD light curves and TESS light curves by Exofast-v1. Also, we collected the mid-transit times published in previous literature on this exoplanet, and in this study, we used unpublished data observed by the TRAPPIST and EulerCam projects. These light curves were given by these projects for use in this study. Figure 11 shows a TTV-diagram of all the mentioned data. We used the linear ephemeris reported by Smith et al. (2012) as a reference ephemeris for computing the epochs and the TTV values.

Reference ephemeris (BJD$_{TDB}$) = $(2455569.83731 \pm 0.00009) + (1.5373653 \pm 0.0000026) \times E$ (13)

Then we derived the new ephemeris by a linear fit of the TTV diagram as

New Ephemeris (BJD$_{TDB}$) = $(2455569.837938 \pm 0.000016) + (1.53736567 \pm 0.00000003) \times E$ (14)

The corner plot of dp and dT obtained from MCMC (50 walkers, 10000 step numbers, and 500 burn-in) is presented in Figure 12. Since there is no obvious parabolic trend, the TTV changes were likely due to wrong linear ephemeris that increased with the number of transit epochs over ten years. However, WASP-36 b requires further observations.

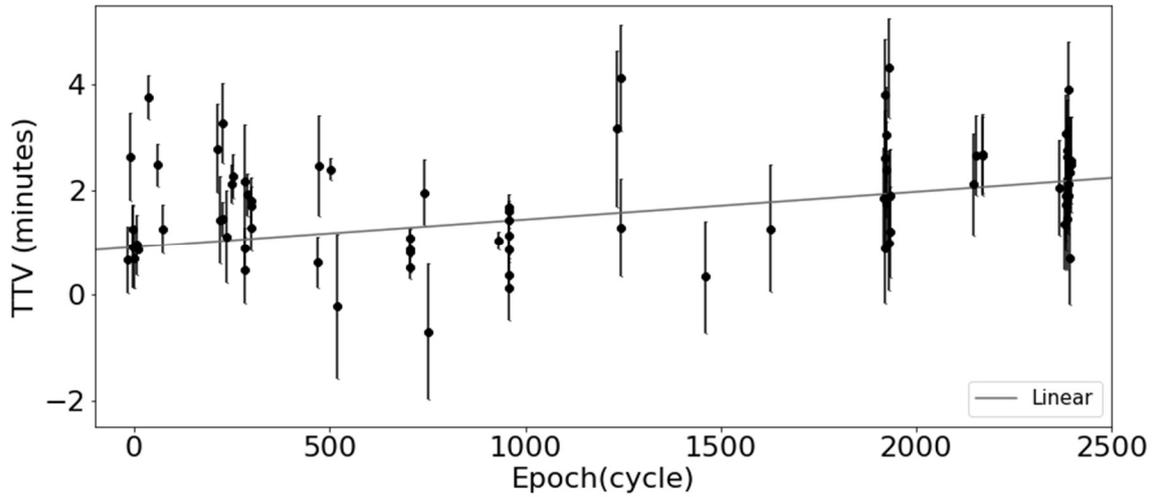

**Fig 11.** The TTV Diagram of WASP-36 b with the linear fit on the data points using the MCMC.

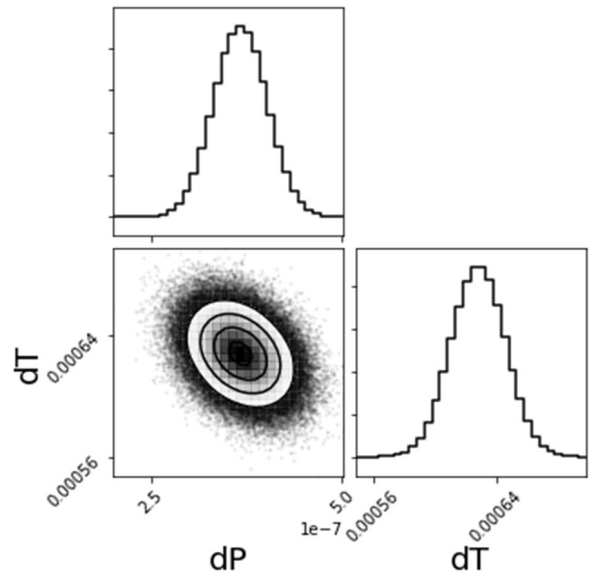

**Fig 12.** Posterior distributions for the fitted parameters using MCMC (dT & dP).

**4.4 WASP-33 b**

WASP-33 b (HD 15082b) is a hot Jupiter discovered in 2010 by the transit method (Cameron et al. 2010). It orbits a δ Scuti variable host star with $V$ = 8.3 magnitude, about 1.5 times the Sun's mass and radius (Herrero et al. 2011). This planet and its host star have been widely studied through spectroscopy and photometry, specially by pulsation studies. However, the first TTV diagram of this planet was analyzed from pre-discovery Hipparcos data by McDonald & Kerins (2018). They calculated the system's new ephemeris. Strong orbital decay for WASP-33 b is less likely, because the star is relatively warm with thin convective envelopes (McDonald & Kerins 2018). Due to its spin-orbit misalignment (Cameron et al. 2010), non-radial changes in the orbit of WASP-33 b may also be expected (Iorio 2011; Lin & Ogilvie 2017). For WASP-33 b, $\delta P/P < -1 \times 10^{-10}$ implies $Q' > 2 \times 10^5$, which is not very limiting, but interesting given the visible tides the planet generates on its star (von Essen et al. 2014). Lehman et al. (2015) examined high-resolution spectra of this system to investigate the effect of the planet. The RV analysis revealed ten frequencies that show the largest amplitude in agreement with the period of the planetary orbit. The contribution of this frequency to the RV variations revealed that it fits the expected photometric orbital solution and planet effect confirmed in RVs. von Essen et al. (2020) updated WASP-33 b ephemeris according to 23 days of TESS data in sector 18, derived from their transit fitting accounting for stellar pulsations. We used the high-quality mid-transit times published in previous literature, and the mid-transit times resulted from the modeling of ETD light curves and TESS for plotting a new TTV diagram. We used the linear ephemeris of Cameron et al. (2010) for computing epochs and the TTV values.

Reference ephemeris (BJD$_{TDB}$) = $(2454163.22373 \pm 0.00026) + (1.2198669 \pm 0.0000012) \times E$    (15)

The TTV diagram of the WASP-33 b is shown in Figure 13. We refined its ephemeris as

New Ephemeris (BJD$_{TDB}$) = $(2454163.224606 ^{+0.000297}_{-0.000298}) + (1.2198704 \pm 0.0000001) \times E$    (16)

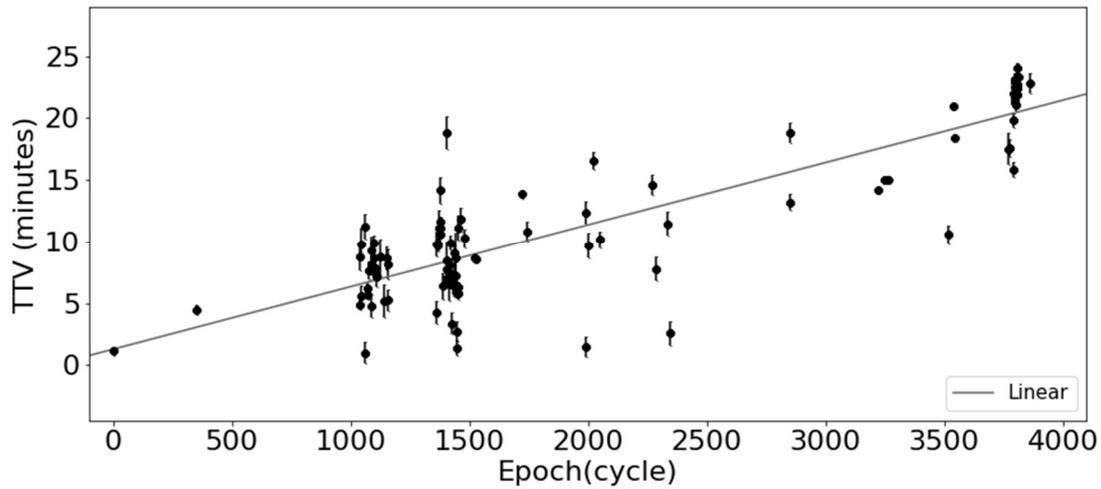

**Fig 13.** The TTV Diagram of WASP-33 b with the linear fit on the data points using the MCMC.

The corner plot of dP and dT obtained from MCMC (50 walkers, 10000 step numbers, and 500 burn-in) is shown in Figure 14. This TTV diagram shows no quadratic trend caused by the effect of the planet's tides on its host star.

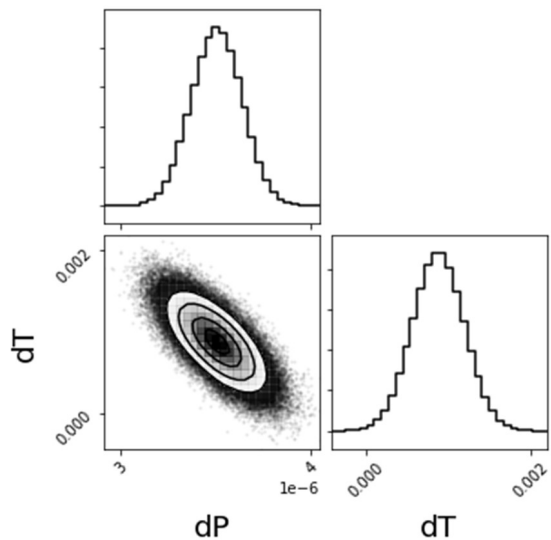

**Fig 14.** Posterior distributions for the fitted parameters using MCMC (dT & dP).

## 5. SUMMARY AND CONCLUSION

We plotted the TTV diagram for WASP-12 b, WASP-33 b, WASP-36 b, and WASP-46 b. The aim is to refine the reference ephemeris of these planetary systems and to discuss the causes of their period changes for future studies. We used the light curves observed by TESS, ground-based telescopes, the TRAPPIST team, and EulerCam observations for WASP-36 b; we applied the mid-transit times obtained in the previous literature too. We modeled the available light curves to extract the mid-transit times using Exofast-v1.

We plotted a new TTV diagram for four systems and refined their ephemeris using the TTV diagrams through the MCMC method. We found a rate of $\frac{dP}{dt} = -25.34 \pm 1.41 \frac{ms}{yr}$ for WASP-12 b, and we derived $Q'_* = (2.13 \pm 0.29) \times 10^5$ for the tidal quality parameter of its host star. This value is also in agreement with $Q'_* \sim 2 \times 10^5$ (Baluev et al. 2019, Patra et al. 2017). It is likely that for WASP-12 b, radial velocity data can bring remarkable information about the nature of this TTV trend (Baluev et al. 2019). We realized that a decrease in orbital period does not continue in WASP-46 b TTV values, indicating that the orbital decay model is weak in this system. The variation in transit times is most likely caused by the magnetic activity of the host star. Since there is no obvious quadratic trend in its TTV diagram, the variations of WASP-36 b are probably caused by the wrong linear ephemeris. As for WASP-33 b, more observations are needed to determine the presence or absence of a tidal effect in this system.

**ACKNOWLEDGMENTS**

This manuscript was prepared to cooperate with the International Occultation Timing Association Middle East section (IOTA/ME) and Tafresh University, Tafresh, Iran. This group activity happened during the Eighth Summer School of Astronomy, held between 21-26 August 2020. We give our special thanks to Dr. Ozgur Basturk for his scientific cooperation. Furthermore, thanks to Dr. Somayeh Khakpash for making some corrections to the text. Authors thanks Tafresh University for the great support of project No. 11/903 that made doing this scientific activity possible.

This study was made possible by the scientific cooperation of the Astronomical Society of the Czech Republic (ETD section) and observational information on its website (http://var2.astro.cz). The Czech Astronomical Society (found in 1924) is an organization coordinating research and observing variable stars and exoplanets in the Czech Republic. Members of this organization are mostly advanced amateur and professional astronomers. We thank the observers (S. Irwin, R. Zambelli, R. Naves, P. Cagas, A. Ayiomamitis, A. Marchini, K. Ivanov, S. Shadic, D. Molina, M. Bretton, N. Esseiva, M. Vrašťák, W. Kang, C. Arena, F. Campos, V. Perroud, Y. Jongen, G. Montanari, V. Hentunen, T. Mollier, T. Scarmato, E. Herrero, L. Brát, G. Corfini, J. Gaitan, J. Lopesino, F. Garcia, B. Guvenen, N. Sebastian, P. Benni, J.L. Salto, M. Raetz, A. Wunsche, G. Marino, P. Pintr, Tan TG, F. Emering, M. Zibar, J. Gonzalez, P. Herbert, A. Carreño, F. Scaggiante, D. Zardin, P. Guerra, S. Ferratfiat, I. Curtis, T. Sauer, C. Colazo, R. Melia, M. Schneiter, E. Fernández-Lajús, P. Romina, Di Sisto, M. Mašek, K. Hoňková, J. Juryšek, C. Villarreal, C. Quiñones, P. Evans, CAAT - Centro Astronomico del Alto Turia).

Furthermore, the assistance provided by the TRAPPIST team and EulerCam includes Michael Gillon (University of Liege, Belgium), Laetitia Delrez (University of Liege, Belgium), Monika Lendl (University of Geneva, Switzerland), Emmanuel Jehin (University of Liege, Belgium), Pierre Magain (University of Liege, Belgium), Cyrielle Opitom (University of Edinburgh, UK) was greatly appreciated for their scientific cooperation in this project.

**APPENDIX**

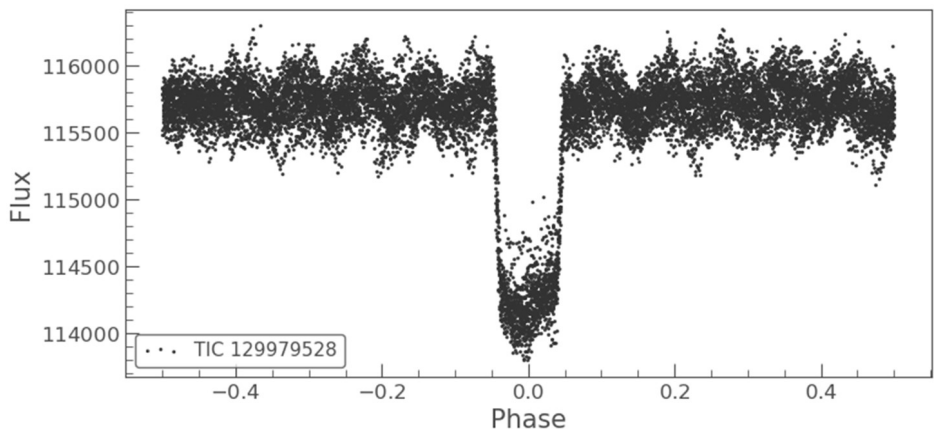

**Fig A.1** The folded light curves of WASP-33 b - Sector 18 (TIC 129979528).

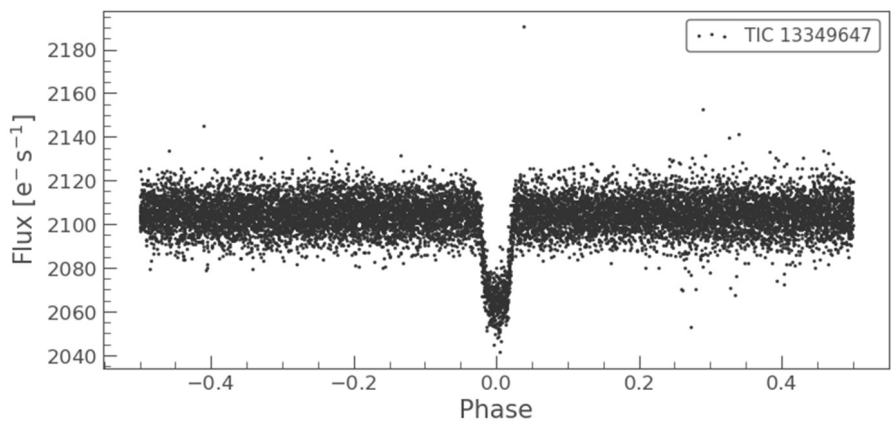

**Fig A.2** The folded light curves of WASP-36 b - Sector 8 (TIC 13349647).

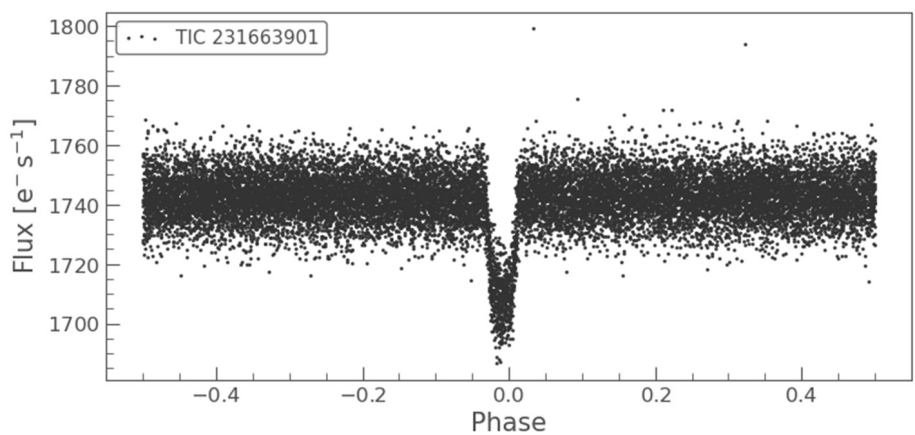

**Fig A.3** The folded light curves of WASP-46 b - Sector 1 (TIC 231663901).

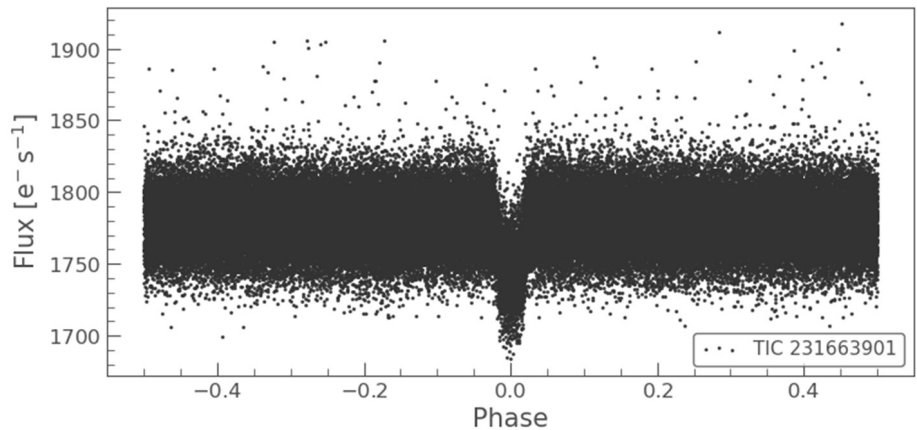

**Fig A.4** The folded light curves of WASP-46 b - Sector 27 (TIC 231663901).